\documentclass[preprint]{aastex}
\usepackage{fullpage}
\usepackage{color}

\usepackage{natbib}

\slugcomment{Not to appear in Nonlearned J., 45.}
\shorttitle{Monte Carlo Simulation with Bulk Diffusion}
\shortauthors{Chang \& Herbst}

\begin{document}

\title{Interstellar Simulations Using A Unified Microscopic-Macroscopic Monte Carlo Model with a full Gas-Grain Network including Bulk Diffusion in Ice Mantles }
\author{Qiang Chang}
\affil{Department of Chemistry, University of Virginia,
              Charlottesville, VA 22904 USA }
\affil{Xinjiang Astronomical Observatory, Chinese Academy of Sciences, 150 Science 1-Street, Urumqi 830011, PR China}
\affil{Key Laboratory of Radio Astronomy, Chinese Academy of Sciences, 2 West Beijing Road, Nanjing 210008, PR China}
\author{Eric Herbst}
\affil{ Departments of Chemistry, Astronomy and Physics,
University of Virginia, Charlottesville, VA 22904 USA}

\begin{abstract}
We have  designed an improved algorithm that enables us to simulate the chemistry of cold dense interstellar clouds with a full gas-grain reaction network.  The chemistry is treated by a unified microscopic-macroscopic 
Monte Carlo approach that includes photon penetration and bulk diffusion.  To determine the significance of these two processes, we simulate the chemistry with three different models.  
 In Model 1, we use an exponential treatment to follow how photons penetrate and 
photodissociate ice species throughout the grain mantle.  Moreover, the products of photodissociation are allowed to diffuse via bulk diffusion and react within the ice mantle. Model 2 is similar to Model 1 but with a slower bulk diffusion rate.  A reference Model 0, which 
only allows photodissociation reactions to occur on the top two layers, is also simulated.
Photodesorption is assumed to occur from the top two layers in all three models.
We found that the abundances of major stable species in grain mantles do not differ much among these three models,
and the results of our simulation for the abundances of these species agree well with observations.  Likewise, the abundances of
gas-phase species in the three models do not vary.
However, the abundances of radicals in grain mantles 
can differ by up to two orders of magnitude depending upon the degree of photon penetration and the bulk diffusion of photodissociation products.  
We also found that complex molecules can be formed at temperatures as low as 10 K in all three models.
\end{abstract}

\section{Introduction}
In cold dense regions of the ISM,  molecules are found in both the gaseous and solid phases. 
Most species including the most abundant one, H$_2$, are found only in the gas phase. 
Some species are found to have appreciable abundances in the layers of ice mantles, which
cover  interstellar grains with possibly more than 100 monolayers of ice in cold dense sources.
The most abundant species in the ice mantle is water,   
the fractional abundance of which with respect to H$_2$ is about $5\times10^{-5}$~\citep{Gibb2004, Pontoppidan2004,Boogert2004,Oberg2011}. The next 
most abundant species are carbon dioxide and carbon monoxide, with abundances   10-50  percent that of water~\citep{Oberg2011}.
Methane, methanol and ammonia possess smaller abundances in the range 1-10 percent of water ~\citep{Oberg2011}. 
Moreover,  there might also be small amounts of other stable 
molecules such as H$_2$O$_2$, the upper limit of which has been investigated recently \citep{Smith2011}.
Some species in the ice mantle, such as carbon monoxide, are formed initially in the gas phase and then accrete onto dust grains while most species are formed from simpler species on the grain surfaces and possibly inside the grain mantles. 
For example, only a small abundance of methanol can  be produced in the gas phase; its dominant mode of production is 
hydrogenation of CO on grain surfaces via reactions with atomic hydrogen, followed by non-thermal desorption \citep{Watanabe2002}.   More complex organic molecules (COMs) such as methyl formate are also believed to be formed on dust grain surfaces  via radical-radical reactions
during the early stages of a later warm-up phase and then released into the gas ~\citep{Garrod2006}.  
Such COMs have recently been detected in the gas-phase of cold dense clouds ~\citep{Bacmann2012}; the details of their formation in these regions are currently not fully understood, but involve both non-thermal desorption of simpler precursor species such as methanol from the grain surface and additional gas-phase reactions~\citep{Vasyunin2013a}.


The treatment of  surface chemical processes on small interstellar grains  involves a number of difficulties.  \citet{Pickles1977} and \citet{Hasegawa1992} modeled
the surface chemistry in the same way as the approach for gas phase chemistry, via coupled chemical kinetic
differential equations. The approach is relatively easy to implement and computationally efficient, thus it is still widely used in gas-grain 
astrochemical modeling, with the additional approximation that all species in all monolayers of the ice are treated in the same manner.  
However, this approach has at least three major problems. The first  occurs when the average number of reactive species such as atomic hydrogen on grain surfaces is  less than one, and significant 
errors may occur unless the discrete numbers of species and their fluctuations are treated in some manner~\citep{Tielens1997,Caselli1998,Herbst2003,Biham2003}. A number of approaches have been suggested to solve this
problem.  In order of increasing computational complexity, these include the modified rate treatment~\citep{Caselli1998,Garrod2008},
the moment equation approach~\citep{Biham2003}, the chemical master equation approach~\citep{Biham2001, Green2001}, 
macroscopic Monte Carlo simulations~\citep{Charnley1998},  and microscopic Monte Carlo simulations~\citep{Chang2005}.  Although all of these approaches have been  applied to simulate the grain mantle chemistry with either a full surface reaction network 
or at least more than 10 reactions on grain surfaces~\citep{Garrod2008,Du2011,Stantcheva2004,Vasyunin2009,Charnley2009,Chang2012}, 
the computational cost becomes more expensive as the approach becomes more rigorous.
 The second problem was first considered by \citet{Hasegawa1993}, who
 argued that only species on the top layer of a grain mantle will react with each other, so that the bulk layers underneath this layer should remain chemically inert as an ice mantle
gradually forms.   This approach, now referred to as a three-phase model (top layer, bulk, gas) was virtually ignored for some time after its initial suggestion.
 Separating species in the bulk from those on the top layer allows abundant and somewhat reactive species such CO or H$_2$CO or even some very reactive radicals  to remain non-reactive in the ice bulk for a long period of time.

 Despite the fact that the microscopic Monte Carlo method can account for the non-reactivity of bulk species
 due to its spatial resolution of species~\citep{Chang2007,Chang2012}, 
different simulation approaches with less computational cost, but which typically do not have the same degree of spatial resolution capability, 
have also been applied to the problem recently~\citep{Garrod2011,Taquet2012,Vasyunin2013b} .  
The third problem with the conventional rate-equation method is that the same 
species on the outermost layer may have different binding energies and diffusion barriers on grain surfaces 
because of surface roughness, chemisorption, or porosity~\citep{Cuppen2005}. When reactive species are very volatile, modeling binding sites on a surface as uniform indistinguishable
sites can result in large errors because stronger binding sites can help some species to be stabilized on grains until higher temperatures are reached,  thus significantly changing the chemical
kinetics. The formation of  H$_2$ from  two atoms  is an example where  stronger binding sites help to increase H$_2$ formation 
at higher temperatures when all of the atomic hydrogen on grain surfaces would otherwise sublime back to the gas before  
combining to form H$_2$~\citep{Cuppen2005,Chang2005}. 
This problem has been solved by improvements in the conventional rate equation approach~\citep{Taquet2012} 
or more rigorous microscopic Monte Carlo simulations~\citep{Cuppen2005,Chang2005}.

It is clearly true that in order for more accurate astrochemical modeling of ice mantles, any chemical kinetics that occurs in the bulk should be different from that on the top  layer, or top several layers of a rough surface. 
However, it is not as true to treat the bulk as completely inert for several reasons.  First of all, volatile species may gradually diffuse out of the mantle and 
then sublime into the gas phase. This effect has been studied by a rate equation approach with the so-called modified three phase model ~\citep{Fayolle2011}. 
In this
model, species are able to swap positions with other species in the bulk so that they are mobile.  Secondly, 
recent calculations show that the probability of UV photons to be absorbed by one monolayer
of water ice is about 0.7\%~\citep{Andersson2008}. Thus, UV photons can easily penetrate through as much as 150 monolayers of species of water ice in grain mantles
and photodissociate species within these monolayers.
So species within the bulk are only partially shielded from UV radiation and those more deeply buried within the bulk
are less likely to undergo photodissociation.  Moreover, the products of photodissociation
on the surface and those within the bulk were found to behave very differently~\citep{Andersson2008}.  In particular, the photodissociated
species within the bulk were found to diffuse and recombine within the bulk while a large fraction of 
those on the surface were found  to directly sublime.
Recent photodissociation laboratory experiments also qualitatively verify that species photodissociated from bulk water ice can diffuse and 
then recombine~\citep{Oberg2009a}.

Motivated by our recent success in unifying a microscopic Monte Carlo simulation for grain chemistry with a macroscopic Monte Carlo simulation 
for the gas phase~\citep{Chang2012},
we modified our microscopic Monte Carlo method for surface chemical reactions in order to include 
reactions and diffusion within the bulk.  Moreover, for the first time,
we included about 300 surface reactions in the surface reaction network, a large number which is feasible because of our improved simulation algorithm. 
Photodissociation reactions are also included and the products of photodissociation are allowed to diffuse within the bulk ice mantle.  
Another recent attempt has been made to include bulk diffusion in an interstellar model.  In this attempt, a three-phase model 
of gas-grain chemistry was utilized with modified rate kinetics  to study the formation of glycine and other COMs 
during the formation of a hot core at 200 K~\citep{Garrod2013a}.   Here,  photodissociation was assumed
to occur at equal rates in all layers, while swapping was assumed to be the mechanism for bulk diffusion. Swapping, however,  may not be the major mechanism for bulk diffusion at low temperatures because of possibly large barriers against the process.  Not surprisingly, \citet{Garrod2013a}
 found that bulk diffusion is not important for the formation of the ice mantle in the early, low temperature, phase of the calculation. 
 
The remainder of this paper is organized as follows. Our bulk and surface model, reaction network, microscopic-macroscopic Monte Carlo approach, and its application to surface and bulk kinetics are described in the next section.  We present our simulation results for cold dense interstellar clouds in Section 3.
In Section 4, we compare these results with those of other models and observational results. Finally, our conclusions and a discussion appear in Section 5.

\section{Physical Model and Monte Carlo Methods}

\subsection{Surface and Bulk Model}

There are two mechanisms for bulk diffusion: substitutional diffusion and interstitial diffusion, although motion in internal pores can possibly be considered a third \citep{Kalvans2013}. 
The bulk diffusion mechanism used in previous models can be classified as
substitutional diffusion~\citep{Garrod2013a,Fayolle2011}, in which pairs of bulk species  have to replace each other in order to diffuse.
 Interstitial diffusion, which refers to diffusion involving sites in between normal sites of the normal lattice structure, 
can be  energetically more favorable because species in interstitial sites are generally less strongly bound to
other species and there is no additional barrier due to swapping in order to diffuse. The diffusion of photodissociated species in the molecular dynamics
simulation  of ~\citet{Andersson2008} can be classified as interstitial
in nature.  In this work, we focus on interstitial diffusion because our model is for 
temperatures below 20 K; thus,  substitutional diffusion is not the major mechanism because of its high barrier. 
On the other hand, because atomic H is found to be able to react with species in the top four monolayers of the ice mantle
by substitutional diffusion~\citep{Cuppen2009}, we include substitutional diffusion in our simulations using a simplified approach
as explained later.

To include interstitial diffusion and bulk reactions during the growth of an ice mantle, we modified our procedure for microscopic Monte Carlo simulation with a model that simulates impurity diffusion in semiconductors~\citep{Akiyama1987}.  As in our earlier work~\citep{Chang2012}, 
grains are assumed to be uniform in size with  radius $r_{\rm d}= 0.1 \mu$m. 
We start from a bare flat surface where the desorption energy 
might be different from that on flat water ice; however, because the initial desorption energy does not affect our results~\citep{Chang2007},
we still assume that it is the same as that on flat ice.
Using a density of regular sites obtained from experiments on high-density ice of 
$1\times 10^{15}$ cm$^{-3}$~\citep{Jenniskens1995}, there are about $10^6$ regular sites on each grain. 
Based on our microscopic Monte Carlo simulation of H$_2$ formation~\citep{Chang2005}, 
a flat surface with $10^4$ sites is large enough to avoid a finite size effect while an 
inhomogeneous (rough) surface can be much smaller to avoid the effect because diffusing 
species such as H can be trapped on the sites with stronger binding strength and so lie in smaller clusters.
A qualitative analysis of sites visited by species on grain surfaces 
shows that increasing the chemical complexity does not increase the minimum surface area needed to avoid
a finite size effect.  The more complex species, the greater the chance that traveling H atoms will react within a smaller cluster of sites.
So we are able to use a surface, which is initially flat and then becomes rough as time evolves,
with only $10^4$  sites, which constitute1\% of the whole surface area in our simulation. 
Based on our earlier work~\citep{Chang2012}, 1\% of gas phase species around one dust particle is also enough to avoid an analogous finite size effect in the gas phase.
We put these $10^4$ sites on a square lattice with dimension 200 $\times$ 200.  
Nodes in the lattice that have both even numbered x and y coordinates represent 
real binding sites on the grain surface, which we term ``normal.''
Each normal site has six  nearest neighbor sites, four  in the same layer and two in
upper and lower layers respectively.  ``Normal'' species, which are located on normal sites, can only move horizontally 
from one normal site to another one by moving two steps, 
 over a potential that includes an intermediate site, which cannot be occupied.
Nodes that have both odd numbered x and y coordinates represent interstitial sites, which are also surrounded by six nearest neighbor sites, four of which lie in the same layer.  Nodes
that have an even numbered and an odd numbered coordinate are forbidden to be occupied at any time.  

\begin{figure}
\centering
\resizebox{8cm}{8cm}{\includegraphics{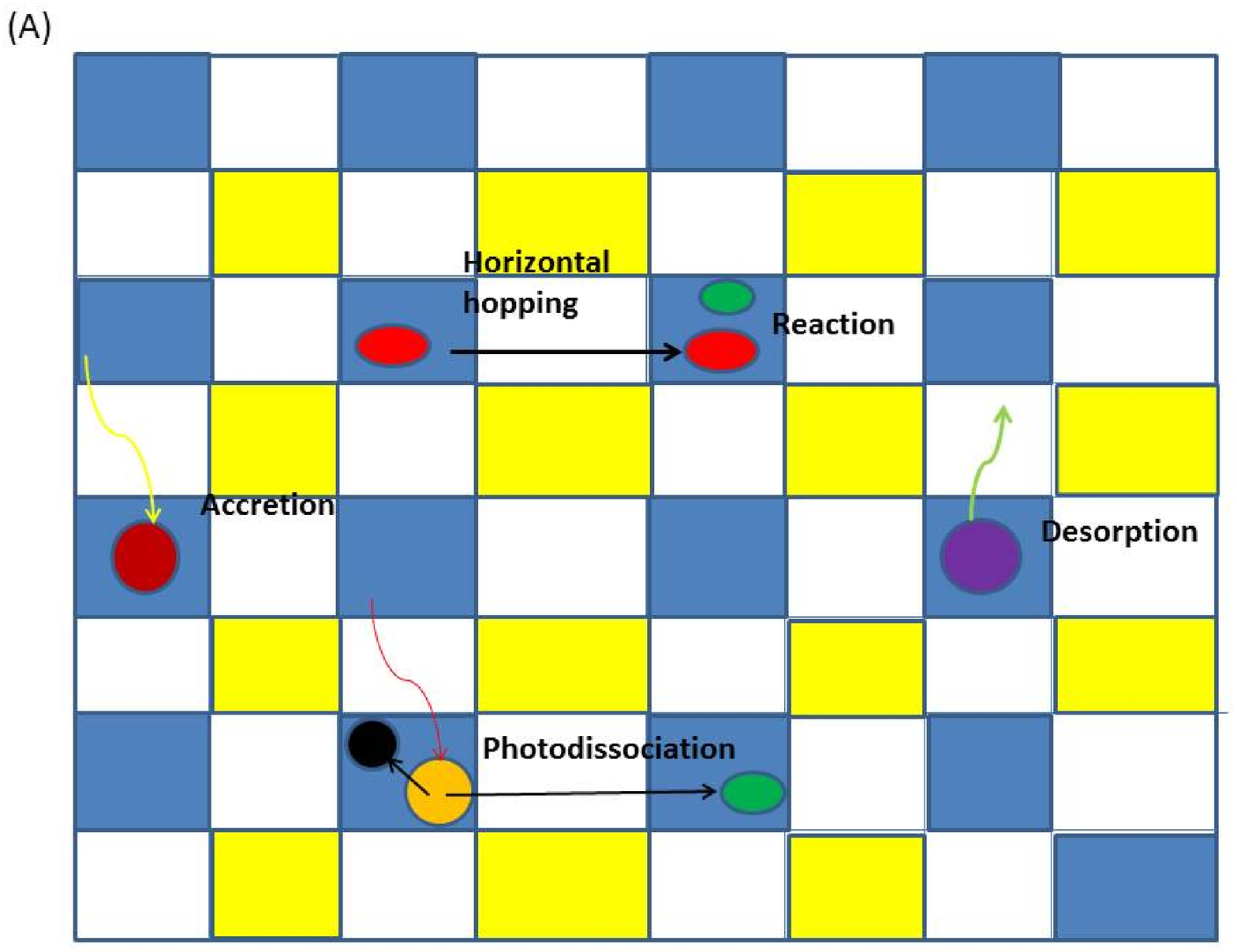}}
\resizebox{8cm}{8cm}{\includegraphics{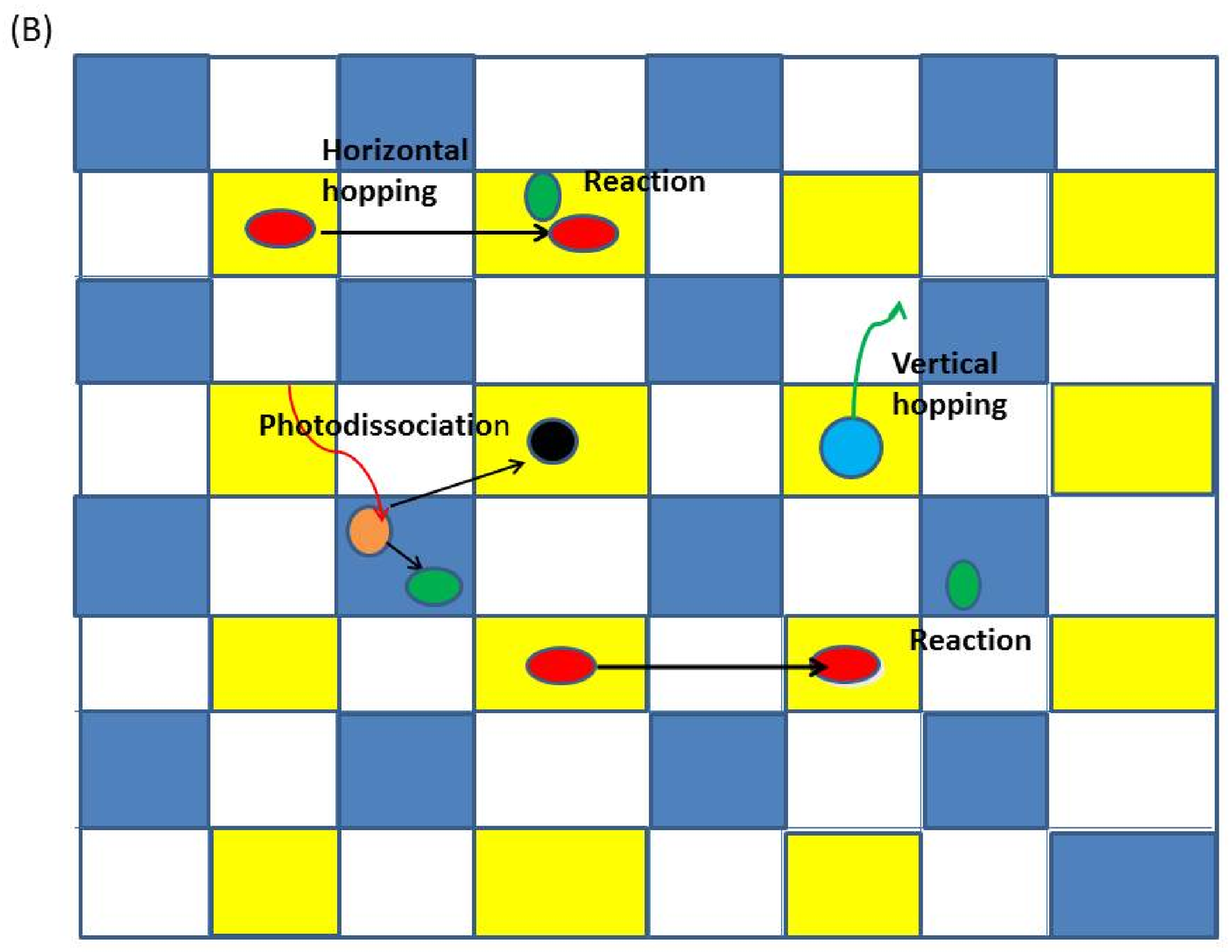}}
\caption{Major processes on grain surfaces.  Blue sites are normal sites, occupied by normal species, while yellow 
sites are interstitial sites, occupied by interstitial species, and white sites cannot be occupied.
Panel (A) shows the topmost layer, while Panel (B) shows an inner (bulk) layer.  The differing processes 
that occur in the topmost layer and the bulk layers are discussed in detail in the text.  Note that although Panel A shows yellow squares, no interstitial sites exist on the topmost layer.
}
\label{fig6}
\end{figure}

The horizontal structure of a monolayer is shown in Panels A and B of Figure~\ref{fig6}.  In these panels, the blue squares represent normal sites, which can only be occupied by normal species, while the yellow squares represent interstitial sites, which can only be occupied by interstitial species.  The white squares cannot be occupied.  What differentiates Panels A and B are the allowed processes on the surface, or topmost, monolayer (Panel A) and any bulk ice monolayer (Panel B).  On the surface, only normal species are allowed.  They hop from one blue square to another, accrete onto blue squares, desorb into the gas from blue squares, and undergo photodissociation from blue squares.  The lighter of the two photodissociation products moves to a nearest neighbor normal site while the heavier of the two remains in the site occupied by the parent.   Finally, chemical reactions can occur when two species are in the same normal site.  These processes mimic the surface processes included in earlier treatments.
Motion in bulk layers involves the interstitial sites: like normal species on the surface, interstitial species in the bulk can undergo horizontal 
hopping to nearest neighbor interstitial sites and undergo reaction if two species wind up on the same interstitial site.  
Interstitial species, however, can also undergo vertical hopping to upper and lower levels, and react if two end up in the same site. 
Conversion between normal and interstitial species occurs via several processes.  Normal species in bulk layers can undergo photodissociation; 
the lighter product ends up in an interstitial site. Interstitial species become normal species when they move to the top layer, 
where there are no interstitial sites.  Essentially, instead of remaining in an interstitial site on the topmost layer, 
the hopping interstitial species is placed in a neighboring normal site.   
There is also a reactive process in which a species in an interstitial site reacts with one in a nearest neighbor normal site, 
with the product to be put in the normal site if there is only one product.  If there is more than one product,  that with 
the largest diffusion barrier is put in the normal site and all other products 
put in neighbor interstitial sites.

In order to simulate the chemical kinetics on and in grain mantles, four parameters have to be 
specified.  The first  is the diffusion barrier on the top layer, $E_{\rm b}$, 
which determines how fast normal species hop from one normal site to another. The hopping 
rate is calculated as $b_1 = \nu \exp(-E_b/T)$ where $\nu$, the second parameter, is the trial frequency and $T$ is the grain temperature.  
The trial frequency is set to be $2\times 10^{11}$ s$^{-1}$ which is lower than the
standard value $10^{12}$ s$^{-1}$. We use the lower value because the temperature dependent reaction barriers 
of H + CO and H + H$_2$CO are fitted with the lower trial frequency~\citep{Fuchs2009}.
The third is the desorption, or sublimation,
energy, $E_{\rm D}$, which determines how rapidly normal species are thermally desorbed. The rate for sublimation is given by
$b_2 = \nu \exp(-E_D/T)$. The last parameter is the diffusion barrier within the bulk, E$_{b2}$, which determines 
how rapidly interstitial species diffuse. The interstitial diffusion rate is given by $b_3 = \nu \exp(-E_{b2}/T)$.
We choose the relation, $E_{b0} = 0.5E_{D0}$ for a flat icy surface~\citep{Garrod2006},
with desorption energies for assorted species taken from ~\citet{Garrod2006}.
However, the real surface may be rough, so normal species on the top
layers may form lateral bonds with neighboring normal species. The actual diffusion barrier for normal 
species  on these top layers is given by the expression 
$E_{bi}=E_{b0i} + nE_{Li}$ where
$E_{Li}=0.1E_{D0i}$ is the strength of lateral bonds formed between normal species and water (the dominant ice), $i$ refers to a particular species,   and $n$ is the number of lateral bonds.   
The strength of lateral bonds is not well known and has been chosen to ensure sufficient mobility for species and yet be
strong enough to form H$_2$ efficiently~\citep{Cuppen2005}.  
We only consider  lateral bonds
between normal species because the bond formed by normal species and 
interstitial species is weaker and also interstitial species will become normal
species when they diffuse to the top of the grain mantle, so there are very few interstitial species on the top few layers. 
Similarly, we have $E_{Di}=E_{D0i} + nE_{Li}$, which only refers to the normal species.
For a detailed explanation of how we calculate $E_{\rm b}$ and $E_{\rm D}$, 
we refer to our earlier paper~\citep{Chang2012}. In fact,
if we ignore interstitial sites and interstitial species, the bulk and surface models here are the same as what we used before~\citep{Chang2012}.
In the work by~\citet{Garrod2013a}, which focuses on substitutional bulk diffusion, the bulk diffusion barrier is chosen
to be $0.7E_D$. Here we focus on interstitial bulk diffusion, which is a different physical process and highly uncertain.
However, because bulk diffusion is normally slower than diffusion on the surface, we choose the barrier
to be $E_{b2i}=0.7E_{D0i}$ or $E_{b2i}=E_{D0i}$. 
The first value will ensure that interstitial species will diffuse more slowly than normal
species do on the top layers while small interstitial  species such as H can still diffuse rapidly enough to lead to reactions even at 10 K.
The second value essentially allows only  the weakly bound atomic H to diffuse within the bulk.  Even atomic H cannot diffuse within the bulk 
if we assume a higher value of $E_{b2}$, so it is not interesting to set $E_{b2}$ at still higher levels.

\subsection{Gas-grain Model}

The physical conditions in our simulation pertain to cold dense cores. The chosen temperature is 10 K or 15 K and the density of H nuclei, $n_{\rm H}$,  is fixed at
$2\times 10^4$ cm$^{-3}$. The cosmic-ray ionization rate is set at $\zeta_{H_2}=1.3 \times 10^{-17}$ s$^{-1}$.  Reactions, rate coefficients, and species are taken
from~\citet{Vasyunin2013b}, with some exceptions.  As noted previously by other authors \citep{Garrod2009}, 
the sparse nature of the surface network can lead to catalytic pathways, which can eat up much of the atomic hydrogen,
 so that H atoms will be converted to H$_2$ and 
so hinder the formation of species such as water and methanol.  The two reactions  below are excellent examples:

\begin{equation}
\begin{array}{llllllll}
\rm{H}  &  + &    \rm{H}_2\rm{O}_2  &   \rightarrow &    \rm{O}_2\rm{H}  &  + &   \rm{H}_2, \\
\rm{H}  &  + &    \rm{HNO}    &    \rightarrow &  \rm{NO}    &  + & \rm{H}_2.  \\
\end{array}
\end{equation}
Although these reactions have barriers of 1900 K and 500 K respectively,  they appear to start catalytic cycles in which the products associate with H atoms to re-form the reactants unless there are competitive pathways, not considered in our network.  It should be noted  that the second reaction does not appear to be important in laboratory experiments leading to the formation of NH$_2$OH on ice \citep{Congiu2012}, while the first reaction must compete with the more standard products H$_{2}$O + OH.   
In addition, as in~\cite{Vasyunin2013b}, 
we do not allow any H$_{2}$ to remain on grains and so include no surface reactions involving H$_{2}$.
We neglect surface H$_{2}$ because the high abundance of
gaseous H$_2$ and the low desorption
energy  of molecular hydrogen on grain surfaces make 
Monte Carlo simulations that include rapid molecular hydrogen accretion and desorption not feasible currently~\citep{Vasyunin2013a}.  

  Some other granular reactions have reaction barriers and we adopt the values in \citet{Vasyunin2013b} 
except for the barriers pertaining to H + CO and H + H$_2$CO, which are temperature 
dependent based on recent laboratory measurements~\citep{Fuchs2009}.  The reaction barrier at 10 K is found
by assuming that reaction rates for CO + H and H$_2$CO + H at 10 K are the same as that at 12 K.
In total, we include 4402 gas phase reactions other than accretion and 459 gas phase species. The number of  
surface/mantle reactions other than photodissociation and desorption is 238 and there are 193 surface species.  The initial gas-phase abundances are 
listed in Table~\ref{table1}, and are based on a particular assumption of low-metal abundances~\citep{Semenov2010}.  We assume an initial value for atomic hydrogen based on gas-phase models. The dust-to-gas number ratio is fixed at $10^{-12}$.

Photodissociation and photodesorption are included in our simulations. The far-ultraviolet (FUV) flux, $F_{FUV}$,  is given by the expression
\begin{equation}
F_{FUV}=G_0F_0 e^{-\gamma Av} + G_{0}^{'}F_0,
\label{radiation}
\end{equation}
where $G_0=1$ is the external radiation scaling factor, 
$G_{0}^{'}=10^{-4}$ is the scaling factor for cosmic ray induced photons based on the  calculation of \citet{Shen2004},
and $F_0$ is the standard interstellar radiation field with a value of $10^8$ cm$^{-2}$ s$^{-1}$. 
The exponential factor $\gamma$ is set at 2 based on the experimental analysis of \citet{Oberg2007}. 
Because the second term on the right-hand side of  eq.~(\ref{radiation}) is much larger than the first term in dense clouds,
the rate coefficient for photodissociation of a species $i$,  $k_{FUV, i}$,  is calculated based on the second term only 
in our simulation.

When photons hit grain surfaces, 
one important parameter for photodissociation is the probability 
that species in each monolayer absorb the photons.
However, little is known about the probability of photon absorption per monolayer for different species except water, which 
was estimated theoretically to be $P_{0}=0.007$~\citep{Andersson2008}. To obtain the probability of photon absorption per monolayer for a species $i$, 
we assume that this probability is proportional to the photodissociation rate on grain surfaces, and that these 
 rates and products are the same as their gas-phase analogs.
The probability of photodissociation for a molecule, $i$, per monolayer is proportional 
to
the overall photodissociation rate. For a molecule , i, it
is $\sum k_{FUV, mi}$, which is the sum of all rate of product channels $m$.
Thus, we can
estimate the probability that a species $i$ absorbs FUV photons per monolayer as
\begin{equation}
P_{i} = P_{0}\frac{\sum k_{FUV, mi}} {\sum k_{FUV, m0}},
\label{photo}
\end{equation}
where $k_{FUV, m0}$ is 
the  photodissociation rate coefficient for water in the m-th 
product channel~\citep{Semenov2010}.  

\begin{table}
\caption{Initial Gas Phase Abundances}
\label{table1}
\begin{tabular}{lc}
  \hline \hline
  Species & Fractional Abundance w.r.t. $n_{\rm H}$ \\
  \hline
  He       & $9.00 \times 10^{-2}$ \\
  H        & $1 \times 10^{-4}$ \\
  H$_2$    & $5 \times 10^{-1}$ \\
  C$^{+}$  & $1.2 \times 10^{-4}$\\
  N        & $7.6 \times 10^{-5}$\\
  O        & $2.56 \times 10^{-4}$\\
  S$^{+}$  & $8.0 \times 10^{-8}$\\
  Si$^{+}$ & $8.0 \times 10^{-9}$\\
  Na$^{+}$ & $2.0 \times 10^{-9}$\\
  Mg$^{+}$ & $7.0 \times 10^{-9}$\\
  Fe$^{+}$ & $3.0 \times 10^{-9}$\\
  P$^{+}$  & $2.0 \times 10^{-10}$\\
  Cl$^{+}$ & $1.0 \times 10^{-9}$\\
  \hline
\end{tabular}
\end{table}

\clearpage

\subsection{Monte Carlo Methods}     
The macroscopic Monte Carlo method for gas phase chemistry here is the same as that used in our previous work~\citep{Chang2012,Gibson2000}. However, 
because the modified surface model we use in this work is more complex 
than that used previously, we have to modify the microscopic Monte Carlo methods.  
In this subsection, we briefly introduce our microscopic Monte Carlo
method  then emphasize the modifications  needed.  
A thorough discussion of the unified  macroscopic-microscopic approach can be found in~\citet{Chang2012}.


The microscopic Monte Carlo simulation of the movement and reactions of normal species on the grain is essentially  the same as used
previously~\citep{Chang2012}. Normal species can hop from one normal site to another
by moving two steps with rate $b_1$, and they can sublime into the gas phase with a rate $b_2$. 
We determine whether a normal species hops or evaporates by
comparing $b_1/(b_1+b_2)$ with a random number $W$ that is uniformly distributed between 0 and 1. 
 (Unless stated otherwise, all random numbers defined in this work lie between 0 and 1.)
If $W$ is smaller, then the normal species hops, 
otherwise it evaporates. The waiting time between events is calculated as $\tau = -\frac{\rm{ln} Q}{b_1+b_2}$ where $Q$ is another random number.  When species accrete onto grain surfaces from the gas phase, they occupy normal sites and become normal species. Chemical
reactions happen by three mechanisms. The first  is the Langmuir--Hinshelwood mechanism, which occurs when normal species hop to sites where
there are reactive species. The second is the Eley-Rideal mechanism, which occurs when species accrete from the gas phase and immediately collide
with species in the sites where they land. The last bears the name ``Chain'' reactions, which we defined in our previous paper~\citep{Chang2012}.  This mechanism occurs when products of a  surface reaction can subsequently
react with species below them. The most significant example is the production of CO$_2$ at 10 K. 
 We use the competition mechanism 
to determine whether a reaction with an energy  barrier can happen or whether the reactants diffuse away~\citep{HM2008}.    
Binding sites, which are potential minima, are pre-defined in our on-lattice microscopic Monte Carlo simulation, 
thus all species are assumed to be ``point'' particles. So, the size of species is not considered. 
A more rigorous approach, known as the off-lattice microscopic Monte Carlo technique, which calculates
the location of potential minima on the fly, can take into account the size of each species on a grain surface~\citep{Garrod2013b}. However, it is not
our purpose to perform off-lattice microscopic Monte Carlo simulations in this paper.

\begin{figure}
\centering
\resizebox{12cm}{12cm}{\includegraphics{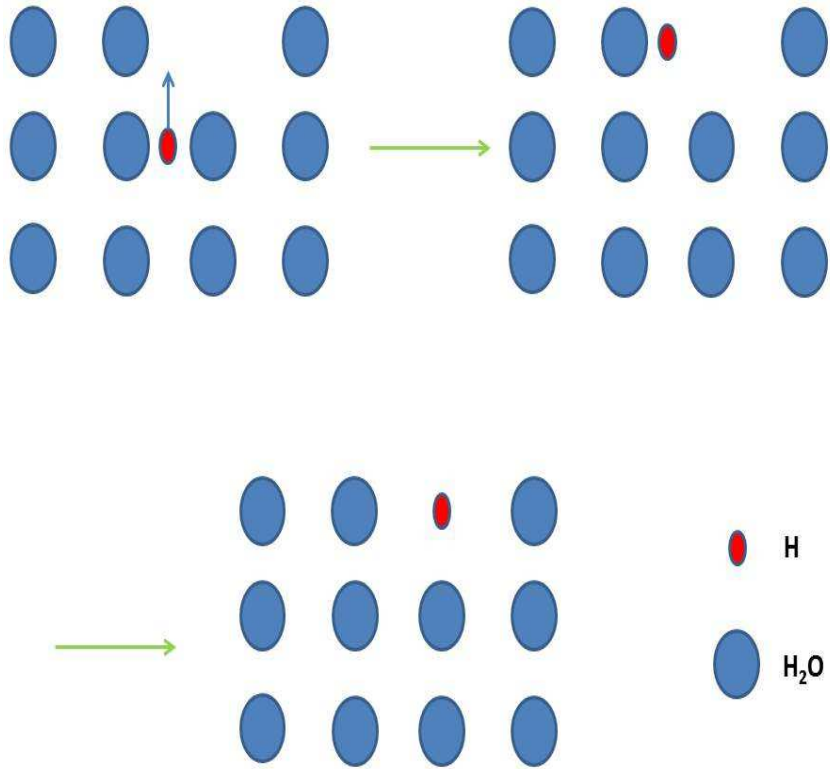}}
\caption{An interstitial H atom, shown in red, undergoes  a series of moves in which it diffuses out of the bulk. The initial movement upwards occurs randomly with a probability of 1/6. See text.
}
\label{figmove}
\end{figure}

The movements and reactions by interstitial species represent the newly added part of the current gas-grain model. Interstitial species can only diffuse from
one interstitial site to another with diffusion rate $b_3$ within the bulk, where the waiting time for diffusion events is $\tau^{'} = -\frac{\rm{ln} X}{b_3}$
with $X$  a random number.
An interstitial species has equal probability to diffuse to its six nearest neighboring sites. 
If the diffusive motion leads to two reactive species  in the same interstitial site, a chemical reaction happens. Moreover, 
whenever an interstitial species comes to a new empty site, or cannot react with a species in the interstitial site, 
we determine if this interstitial species can react with any of its four nearest neighboring normal species
in the same layer. This move is based on the molecular dynamics calculation by \citet{Andersson2008}, in which photodissociation
products can recombine. Thus interstitial species must be able to react with normal species in our model.
If there is more than one neighboring normal species that can react with the interstitial species, we 
randomly pick one normal species to react.
We put the product species with the largest desorption energy 
in the normal site and all other product species are put into neighboring interstitial sites 
around the normal site. 
However, if an interstitial species comes to an already occupied interstitial site and there is no reaction with the interstitial species or  
neighboring normal species 
it is a null event, and the interstitial species is returned to its original site. 
This procedure is undertaken to ensure that an interstitial site is either empty or occupied only by one species.
 In addition, if an
interstitial species randomly comes to an interstitial site with no species right above it (closer to the surface), 
we allow the interstitial species to continue to move, first to proceed to the empty interstitial site immediately above it and then to a neighboring normal site.  This series of moves simulates the process of how volatile species diffuse out of the ice mantle.  
The process is illustrated by the movement of an interstitial H atom in a matrix of normal water molecules, as depicted in Figure \ref{figmove}.

Photodissociation processes are the only ones that change normal species into interstitial ones. The FUV photon arrival rate  onto a grain is calculated as 
\begin{equation}
k_{pho}=F_{FUV}\pi r_{d}^2,
\end{equation}
where $F_{FUV}$ is the FUV flux as in eq.~(\ref{radiation})      
while the waiting time for an FUV photon arrival is calculated as
$\tau^{''} = -\frac{\rm{ln} X^{'}}{k_{pho}}$,
where $X^{'}$ is a random number.
In addition to photodissociation,  photodesorption can also occur when FUV photons arrive on grain surfaces. 
We assume that photodesorption occurs only in the top two monolayers with yields per photon of
$Y_{\rm d }= 0.001$ for all species, a value supported by recent laboratory measurements~\citep{Oberg2009b, Oberg2009c}. 
The Monte Carlo simulation of photon arrival events is done in the following manner. First,  we randomly pick a normal site where an FUV photon will hit. We ignore
photodissociation and photodesorption of interstitial species because they are much less abundant than normal species. 
For a normal species $i$ on the topmost
monolayer, we generate a random number, $Y$. If $Y<Y_{\rm d}$, photodesorption occurs and the normal
species desorbs into the gas phase. If $Y>Y_{\rm d}$ and $Y<P_{i}$ (see eq.(\ref{photo})), the normal species will be photodissociated. 
 In order to
determine the specific channel that photodissociation occurs into, we generate another random number, $Y^{'}$, and compare this number with the branching fractions of products in the specific channels according to the method of~\citet{Gillespie1976}.


  All products of photodissociated species in the topmost monolayer are normal species.
If $Y>P_{i}$, the FUV photon will penetrate through the topmost monolayer 
and we determine if normal species in the monolayer below the topmost monolayer 
can be photodesorbed or photodissociated in a similar manner to what we did for normal species in the topmost monolayer.
The only difference is that if 
photodissociation occurs from the second layer,  the  product
with  the largest 
desorption energy will still occupy the normal site while the products  with smaller desorption energies will occupy neighboring interstitial sites and become interstitial species  if the neighboring normal sites are occupied.  If, on the other hand, the neighboring sites are unoccupied, the products will occupy them. 
If the normal species below the topmost layer can neither be photodissociated nor photodesorbed, 
we move on to the species directly below it, although now we only have to 
check if it can be photodissociated. The products of the photodissociation of species at this monolayer 
are treated in the same way as we did for photodissociation of normal species below the topmost monolayer, 
We continue in this manner until the FUV photon is absorbed or hits the grain base, which is a null event.   
 
The probability that a photon can penetrate through n-1 layers of ice and be absorbed
by the n-th layer of ice is
\begin{equation}
P_{pho_n} = P_{i_n}\Pi_{m=1}^{m=n-1} (1-P_{i_m}).
\end{equation}
where $i_m$ is the species at the m-th layer on the site where photons hit and we assume the topmost 
layer to be the first.
The exponential nature of the penetration is obvious if the ice mantle is composed of pure water ice, so that
\begin{equation}
P_{pho_n} = P_{0}(1-P_{0})^{n-1}.
\end{equation}

We have run three models of cold dense clouds, each at temperatures of 10 K and 15 K.  In Model 1, 
the barrier for interstitial species to move from one interstitial space to another, $E_{b2}$, is set at $0.7E_{D0}$ (see Section 2.1).     Model 2 contains a high barrier $E_{b2} = E_{D0}$.
Because we are also interested in how photon penetration affects ice formation on dust grains, we also run a model, 
labelled Model 0, in which
FUV photons can only photodissociate
species in the top two monolayers, as occurs for photodesorption in all three models.
For Model 0, we also put the products of photodissociation 
in the monolayer below the topmost monolayer into the topmost monolayer in order not to have
any interstitial species in the system, which is comparable to an earlier approach to photodissociation~\citep{Vasyunin2013a}.
Since there are no interstitial species in Model 0, there is no interstitial bulk diffusion although
substitutional bulk diffusion in the first 4 monolayers still exists.

To run a Monte Carlo simulation, we must continually determine the event that will occur first and then execute that event.
In order to do that for our microscopic simulation, we find the time $t_{g}$ of the first chemical reaction to occur in the gas phase, the time $t_{n}$ for  the first hopping or evaporation time 
of normal species in the topmost monolayer  and 
the first time $t_i$ when interstitial species diffuse.
We then find the minimum of these times and  the photon arrival time $t_p$ to define the overall event to occur first: $t_m = min(t_g, t_n, t_i, t_p)$, and then execute the event. The process repeats itself. To find $t_n$ at each time step,
we have to compare up to $L^2$ absolute evaporation or hopping times, where $L^2$ is the total number
of normal sites at each layer.    Species such as atomic H can hop
very quickly, and the time step is very small when fast hopping exists on the grain surface. This problem can make the computational cost very expensive. In order
to solve this problem, we have modified our Monte Carlo methods by dividing normal species into three categories based on the diffusion barrier. 
Atomic H, with the lowest barrier, belongs to
category A; atomic O, C, and N, with somewhat higher barriers,  belong to category B; and all other species belong to category C.  We find
the first hopping or evaporation time $t_{nA}, t_{nB}, t_{nC}$ in each of these three categories, separately. 
We then find the minimum of the three times: $t_n = min(t_{nA}, t_{nB}, t_{nC})$. If, for example, atomic hydrogen hops or evaporates, we only have to
update the absolute time of the first events in category A, which is much faster than comparing all absolute hopping or evaporation times
for normal species on the grain surface because there are typically at most a few H atoms present. Thus, although 
H atoms hop very quickly, the computational cost for each hop is much smaller.  We found that this simple modification can reduce the
simulation time by a factor of 30 at 10 K. Similarly, we divide interstitial species into three categories and find the first
diffusing interstitial species by comparing the absolute time of the first events in each category.   
 This more efficient algorithm is a low-temperature one; as the temperature rises,  the distinction among the three classes is reduced.

\section{Results}   

We perform our simulations under standard dense cloud physical conditions, as described in section 2.2. Each simulation is run for a time
period of $2\times 10^5$ yr, at which time the abundance of water ice on grain surfaces fits observations. Moreover, the simulated abundances of most gas-phase species 
typically agree well with observational results at this time. 
As in our previous paper~\citep{Chang2012}, which only contained a small number of grain surface reactions, each model simulation was run 4 times with different random seeds and the results we report here are
averages of these four simulations.  Because our calculated gas-phase abundances during this time interval do not vary much among our three models and previous work, we focus on the ice mantle species.   We first report results of the major stable granular species 
H$_2$O, CO, CO$_2$, NH$_3$, CH$_4$, H$_2$CO, and CH$_{3}$OH because these species can 
be compared with observations.  Except for ammonia, these species were discussed in our previous paper~\citep{Chang2012}.  
Because radical abundances are sensitive to photon penetration and bulk diffusion, we emphasize the abundances of the radicals OH, CH$_2$OH, and HCO and compare our results with previous work. The radical OH was chosen
because it is one product of the photodissociation of the water ice molecule, which is the most abundant species in ice mantles, 
while CH$_2$OH and HCO are the reactants that combine to form methyl formate at higher temperatures.  We do not differentiate between 
CH$_2$OH and CH$_3$O in our reaction network.

\subsection{Major species}
 
\begin{figure}
\centering
\resizebox{12cm}{12cm}{\includegraphics{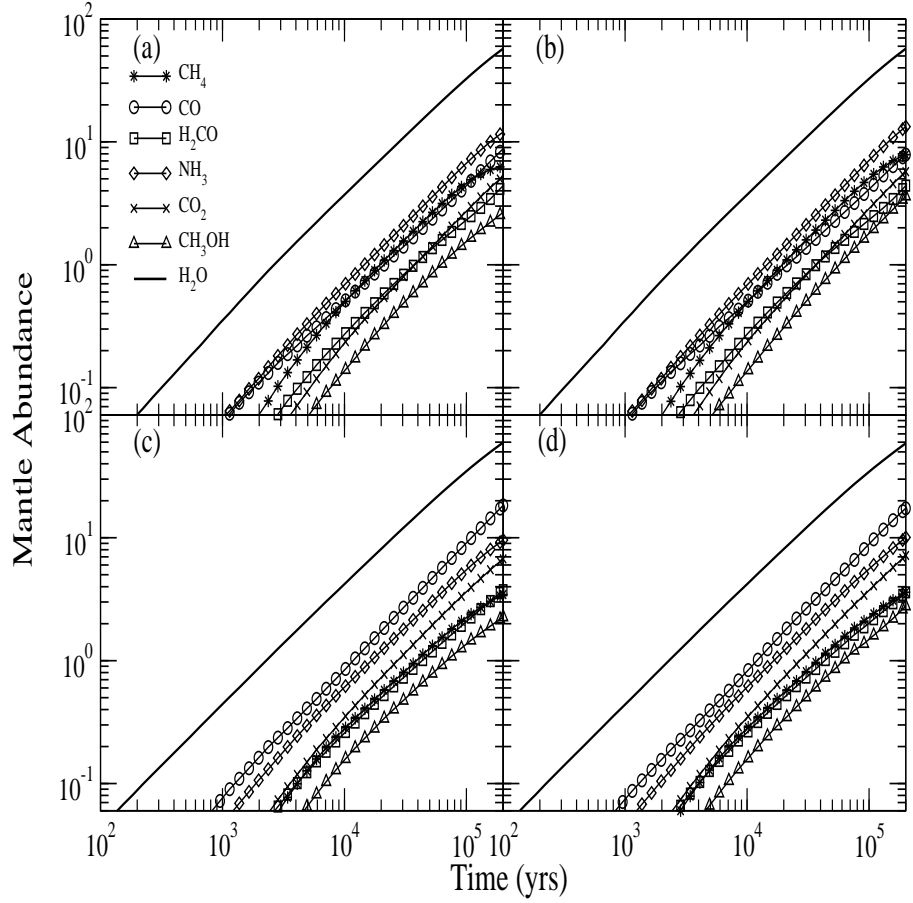}}
\caption{The abundances of major stable species in the ice mantle are shown in terms of monolayers as a function of time.
Panel (a): $T=10$~K, Model 1,
 panel (b): $T=10$~K, Model 0, panel (c): $T=15$~K, Model 1, panel (d): $T=15$~K, Model 0. 
}
\label{fig2}
\end{figure}

Figure~\ref{fig2} shows and compares the abundance of major stable species 
as a function of time at 10 K and 15 K with Model 1 and Model 0
respectively. We do not show results from Model 2 because they are very much like those of Model 1.
The abundance of a species is the sum of the species that occupy normal sites and interstitial sites. It  
is expressed in monolayers; one monolayer of species is equivalent to a fractional abundance of $10^{-6}$  with 
respect to the density of total H atoms. As can be seen, the abundances of all major surface species increase with time.  This increase occurs because although species
such as CO may react with H atoms, they cannot continue reacting efficiently after being buried, 
even if some interstitial species can be formed by photodissociation. 

Comparing the results from Model 1 and Model 0, we can see that photon penetration and bulk diffusion (initiated by photodissociation) hardly change 
the abundance of major stable species, which is not surprising because photodissociation is not a major chemical process in dense clouds. Thus,
the fact that species deeply buried in ice mantles can be photodissociated does not change the total abundances of major species significantly. Moreover, 
as in the gas phase, surface species do not change much when the temperature is increased from 10 K to 15 K. The only difference is that NH$_3$ 
is the second most abundant species at 10 K while CO is the second most abundant species at 15 K, which can be explained by the fact that
the abundance of H atoms decreases at higher temperatures and so cannot hydrogenate nitrogen atoms as efficiently.  We  summarize
the abundances of major stable species for all models at $2\times 10^5$ yr in Table~\ref{table3}.  

\begin{figure}
\centering
\resizebox{10cm}{10cm}{\includegraphics{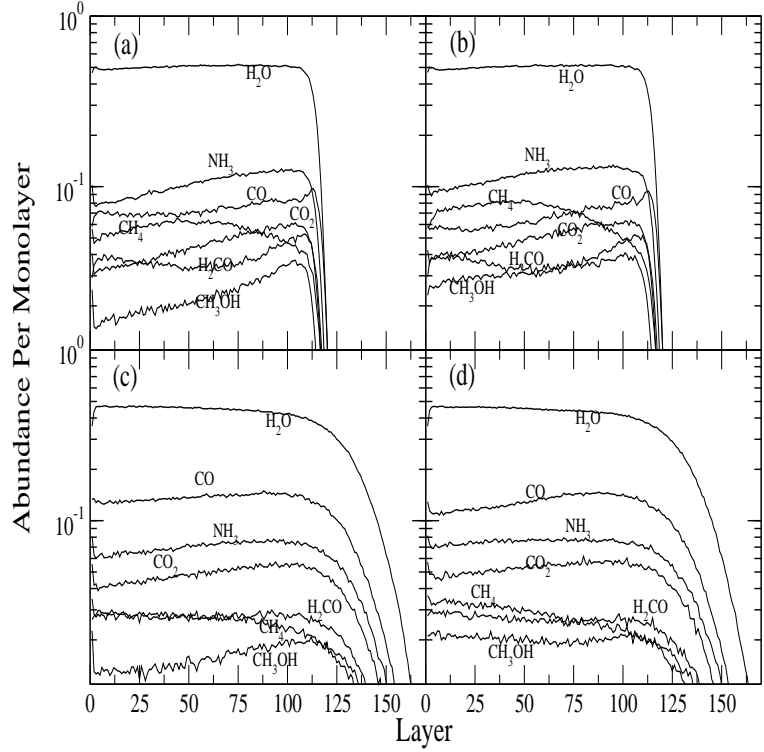}}
\caption{The fraction of each monolayer occupied by major stable species in  Models 1 and 0 shown as a function of monolayer 
at the final time of $2 \times 10^{5}$~yr.
Panel (a): $T=10$~K, Model 1,
 panel (b): $T=10$~K, Model 0, panel (c): $T=15$~K, Model 1, panel (d): $T=15$~K, Model 0. }
\label{fig3}
\end{figure}

Figure~\ref{fig3} shows the fraction of each monolayer occupied by major stable species. We see again that Model 1 and Model 0 results differ 
very little because photodissociation reactions have minor effects on the abundance of major stable species.  Model 2 results
are once again not shown  because they are very similar to those of Model 1. The temperature
difference, on the other hand, can produce a significant difference in the outer layers,
because the mobility of species increases exponentially with temperature.
 Regarding the dependence of abundance on layer,  the abundance
of water almost stays constant,  which shows that water can be efficiently formed on grain surfaces in all models.
The surface CO abundance increases at the outer layers when  
gas phase CO increases. 
The gasous atomic H abundance almost remains constant in the simulation while
gaseous species such as O, which can react with atomic H on grain surfaces without a barrier,   decrease,
so species such as CO can be more easily hydrogenated at the outer layers. Thus,
methanol generally also increases at the outer layers because of the increase in CO  and lessened ability of less abundant species such as O to compete for H atoms. The abundance variation of H$_2$CO is complex because it is an intermediate product of the gradual hydrogenation of CO to methanol.
 Ammonia also increases slightly in abundance at the outer layers because of the greater availability of H atoms despite the gradual decrease of gas phase N.  Carbon
dioxide is mainly produced via the chain mechanism between CO and OH~\citep{Chang2012}, so its abundance increases as the CO abundance increases.

The results concerning the abundances of major species in the ice clearly do not depend much on interstitial sites.   
Table~\ref{table7} shows the abundances of the major
stable interstitial species in Models 1 and 2 as a percentage of the overall 
water ice abundance, and one can see immediately that the numbers are all quite small. 
Overall the abundance of interstitial species is much lower than normal species because photodissociation is the only
process that can convert normal species into interstitial species; however, the radiation level in dense clouds
is so low that photodissociation reactions  occur far less frequently than accretion.  In the following detailed discussion of the interstitial abundances, we use $N(i)$ to represent normal species $i$, 
and $I(i)$ to represent  the interstitial species.
Interstitial CO can be produced by direct photodissociation such as via
N(HNCO) $\rightarrow$ I(CO) + N(NH) or by the association of two interstitial species; e.g.,  I(O) + I(C) $\rightarrow$ I(CO).
We can see that the I(CO) abundance increases as temperature increases in both Model 1 and Model 2. On the other hand,
because I(O) and I(C)  can hardly move even at 15 K in Model 2, the increase of I(CO) mainly comes from the increased photodissociation of
normal species such as N(HNCO). Species
other than I(CO) in Table~\ref{table7} cannot be produced  by direct photodissociation because
few photodissociation reactions have them as products. Moreover, even if these molecules
appear in the products of photodissociation reactions, since their diffusion barriers are typically 
larger than other molecules, they still occupy normal sites in our simulation. These interstitial species in 
Table~\ref{table7} have to be produced by reactions involving solely  interstitial molecules. We can see that as temperature 
increases, their abundances typically increase with temperature 
because of the increased mobility of reactive interstitial species.
It is also interesting to see that despite the fact that N(CO) can be sufficiently hydrogenated all the way to
form N(CH$_3$OH) on the topmost layer, it is not easy to hydrogenate I(CO) sufficiently
to form I(CH$_3$OH) within the bulk although I(H$_2$CO) can be produced by gradual hydrogenation of I(CO).  These facts
 suggest that species occupying interstitial sites within the bulk
are less likely to be hydrogenated than species occupying normal sites on the topmost layer.
Another interesting fact is that I(CO) is the most abundant stable interstitial species instead of water,
because CO is a direct product of photodissociation reactions while water molecules are not.

\begin{table}
\caption{Abundances of Major Stable Mantle Species }
\label{table3}
\begin{tabular}{ccccccc}
  \hline
Species  &      Model 1 & Model 1 & Model 2 & Model 2 & Model 0  & Model 0 \\ 
         &      10 K    & 15 K                &  10 K   & 15 K    & 10 K     & 15 K \\ \hline
CO       &      15.3           & 30.8          & 14.9          & 30.5           & 13.7          &  29.6      \\
H$_2$CO  &      7.7           & 6.3           & 7.4           & 6.1            & 7.7           &  6.1      \\
CO$_2$   &      9.1            & 11.3          & 9.1           & 11.3           & 10.0          &  12.2      \\
CH$_3$OH &      4.7            & 3.8           & 4.7           & 3.8            & 6.3           &  4.7      \\
CH$_4$   &      11.0           & 5.8           & 11.1          & 5.3            & 13.7          &  6.1      \\
NH$_3$   &      20.9           & 16.1          & 20.9          & 15.9           & 23.2          &  17.2      \\
H$_2$O   &      5.7(-5)        & 6.0(-5)       & 5.8(-5)       & 5.9(-5)        & 5.7(-5)       & 5.9(-5) \\ \hline
\end{tabular}
\tablecomments{The abundances refer to a time of $2\times 10^5$~yr for Models 0, 1, and 2 at different temperatures.
The abundance of total water ice(normal + interstitial) is the fractional abundance with respect to  gas-phase $n_{\rm H}$, 
while the abundances of the other ice components are percentages with respect to the total water ice. a(-b) means a$\times 10^{-b}$.
}
\end{table}


\begin{table}
\caption{Abundances of Major Stable Mantle Interstitial Species }
\label{table7}
\begin{tabular}{ccccccc}
  \hline
Species  & Model 1, 10 K  & Model 1, 15 K & Model 2, 10 K & Model 2, 15 K \\ \hline
CO       & 1.3(-1)        & 4.5(-1)       &  1.2(-1)      &  4.6(-1) \\
H$_2$CO  & 1.1(-3)        & 2.4(-3)       &  9.1(-4)      &  2.3(-3)  \\
CO$_2$   & 0              & 2.4(-3)       &  0            &  4.2(-4) \\
CH$_3$OH & 0              & 1.0(-3)       &  0            &  4.2(-4)       \\
CH$_4$   & 2.2(-2)        & 3.7(-2)       &  1.8(-2)      &  1.8(-2) \\
NH$_3$   & 0              & 0             &  0            &  0   \\
H$_2$O   & 1.1(-2)        & 6.1(-2)       &  8.4(-3)      &  9.9(-3) \\
\end{tabular}
\tablecomments{The abundances refer to a time of  $2\times 10^5$~yr for Models 1 and 2 at 
different temperatures.
The abundances are percentages with respect to the overall (normal + interstitial) water ice abundance; thus the CO abundance for Model 1 at 10 K is 1.3(-3) of the water abundance. a(-b) means a$\times 10^{-b}$.
}
\end{table}

\subsection{Radicals}

\begin{figure}
\centering
\resizebox{10cm}{10cm}{\includegraphics{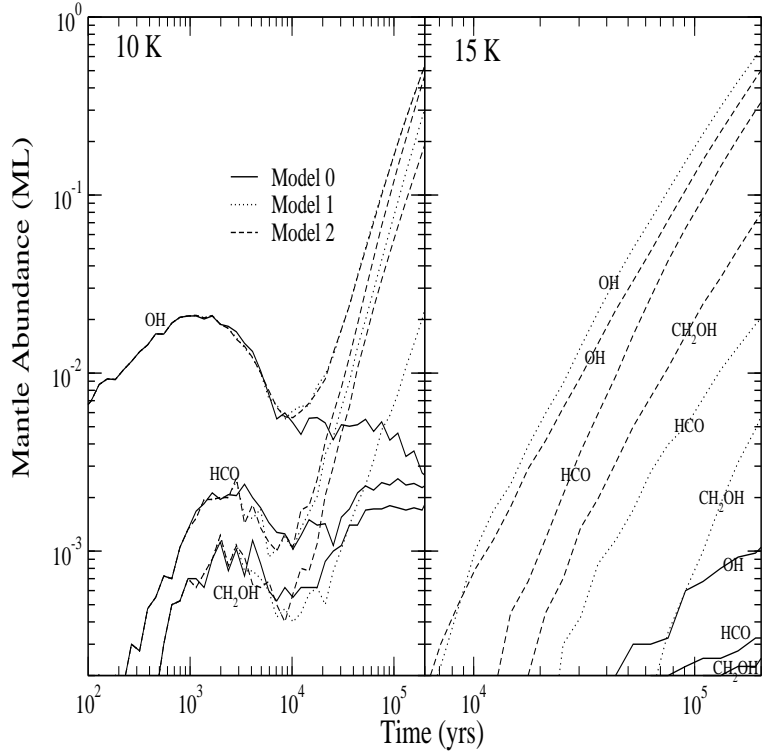}}
\caption{The abundance of selected radicals as a function of time for models 0, 1, and 2.
Panel (a): $T=10$~K,
 panel (b): $T=15$~K. }
\label{fig4}
\end{figure}

Figure~\ref{fig4} shows the abundance (in monolayers) of selected granular radicals as a function of time for Models 0, 1, and 2. 
First, we compare Models 1 and 0 to determine the 
influence of photon penetration and bulk diffusion.
Contrary to the stable major species, the abundances of radicals in Model 1 are about 1 or 2 orders of magnitude higher than in Model 0 at 
$2\times 10^5$ yr, although the abundances of radicals in the three models are similar at earlier times ($<10^4$ yr) at 10 K. 
Before the system has evolved through $10^4$ yr, there are fewer than 4 monolayers of ice formed on the grain surface. 
The H atoms on the mantle surface are able to react with radicals on these monolayers via our H atom swap mechanism, 
 so normal radicals produced before $10^4$ yr are still reactive in all three models. Moreover,
interstitial H, which is the only interstitial species that moves at 10 K, can easily diffuse out of the ice mantle when it
is very thin, so that interstitial chemistry is not important before $10^4$ yr.  So, we cannot see any  difference for
radicals in the three models at these times.    
However, when the system has evolved longer, there are enough monolayers of ice covering the grain surface so that
radicals produced by photodissociation and deeply buried below top 4 monolayers are no longer able to 
react with surface H atoms.
However, interstitial chemistry becomes more important when it is more difficult for bulk atomic H to diffuse out of the ice mantle.  
Thus, the difference between these models quickly becomes larger as time evolves. At 15 K, however, because species diffuse
more quickly,  radicals produced on the topmost layer can easily react with other species before they are deeply buried below the top 4 monolayers
of ice. So we can see that there are few radicals produced before $10^4$ yr.

It is also interesting to see where these radicals are located in the ice mantle in Model 1 because the fraction of radicals in interstitial sites can measure the significance of bulk diffusion compared with direct photodissociation to produce interstitial radicals.  First consider the radical OH.  If it formed mainly from the photodissociation of water, it would stay on the normal site while the lighter H atom would go to an interstitial site.  Thus, OH in an interstitial site is probably formed just by the association between O and H via bulk diffusion.   Now consider the HCO and CH$_2$OH radicals.  If they were formed mainly via photodissociation of the parent species formaldehyde and methanol, they would also remain on normal sites.  
But, almost all HCO and CH$_2$OH are hydrogenation products of CO molecules in interstitial sites, the formation mechanism of which has been
discussed in detail in the last subsection.
Thus, in general, the fraction of radicals
in interstitial sites can measure the significance of bulk diffusion compared with direct photodissociation.
Radicals on normal sites can be formed by photodissociation or bulk diffusion,
so cannot be used to measure the significance of bulk diffusion. 

Table~\ref{table4} shows the fraction of radicals in interstitial sites at $1.1\times10^5$ and $2.0\times 10^5$ yrs.  We can see that
the fractions range from greater than 0.5 to very small values.  The large values are most likely associated with bulk diffusion, as discussed above.  
In order to check if there is any time dependence,  
we show two different times. We can see that generally these fractions do not change much with time.
The dependence of the fraction of interstitial radicals on temperature is very complicated. 
As the temperature increases, the mobility of interstitial species increases, which can both increase
the production of interstitial radicals and eliminate interstitial radicals faster. Moreover, 
we saw that I(CO), which can be hydrogenated to form I(HCO),  increases as temperature increases in the last subsection.
Because of the above competing mechanisms, in Model 1, the fraction of  all radicals in 
interstitial sites increases as the temperature is raised to 15 K,
while, in Model 2, the fraction of CH$_2$OH in these sites increases,  the fraction of HCO 
does not change much,  while the fraction of OH does not change at $1.1 \times 10^5$ yrs and 
slightly increases at $2 \times 10^5$ yrs.

 We can also compare Models 1 and 2 to find out the influence of the barrier difference on the abundance of radicals.
At 10 K, as we explained earlier, the abundances of all radicals
do not depend on the diffusion rate before $10^4$ years. After $10^4$ yr, the abundance of OH is almost the same in both models, 
while the abundances of CH$_2$OH in Model 2 are more than one order of 
magnitude larger than those in Model 1 at both 10 K and 15 K. 
The abundance of HCO in Model 2 is about 2 times as much as that in Model 1 at 10 K while
at 15 K, the HCO abundance is more than one order of magnitude larger than that in Model 1. 
These facts can be explained by the increased mobility of interstitial species in Model 1. 
The influence of the barrier difference on the bulk diffusion rate is also reflected in the fraction of radicals in interstitial sites.
The mobility of species on interstitial sites in Model 2  at 15 K
is approximately the same as that for interstitial species at 10 K in Model 1, so we can see that the fraction in Model 2 at 15 K is very close to 
the fraction at 10 K in Model 1. 

Finally, it is interesting to see which radical is the most abundant. 
The most abundant radicals are temperature dependent. For instance, in both Model 1 and Model 2 at 10 K, the most abundant radicals are
CH$_3$ and NH$_2$, the 
abundances of which are around one monolayer (a fractional abundance with respect to hydrogen of 10$^{-6}$) while at 15 K, NH$_2$,   
the abundance of which is more than two thirds of a monolayer, is the most abundant radical in both Model 1 and Model 2.
  
\begin{table}
\caption{Fraction of Radicals in interstitial sites}
\label{table4}
\begin{tabular}{ccccccc}
  \hline 
Radicals      & OH                &   HCO        & CH$_2$OH  &  OH                &   HCO              & CH$_2$OH \\ 
              & 1.1(5) yr         &  1.1(5) yr   & 1.1(5) yr &  2.0(5) yr         &  2.0(5) yr         & 2.0(5) yr \\ \hline

Model 1, 10 K &  0.40             & 0.10         &  3.3(-3)  &  0.36              &  7.7(-2)           & 1.9(-2)   \\

Model 1, 15 K &  0.58             & 0.82         &  0.50     &  0.63              &  0.83              & 0.45  \\

Model 2, 10 K &  0.36             & 6.9(-2)      &  3.8(-4)  &  0.33              &5.3(-2)             & 1.3(-4) \\

Model 2, 15 K &  0.36             & 6.3(-2)      &  8.2(-3)  &  0.43              & 5.6(-2)             & 1.8(-2)     \\
\hline
\end{tabular}
\tablecomments{a(-b) means a$\times 10^{-b}$.}
\end{table}

\subsection{Complex Organic Molecules}

\begin{figure}
\centering
\resizebox{10cm}{10cm}{\includegraphics{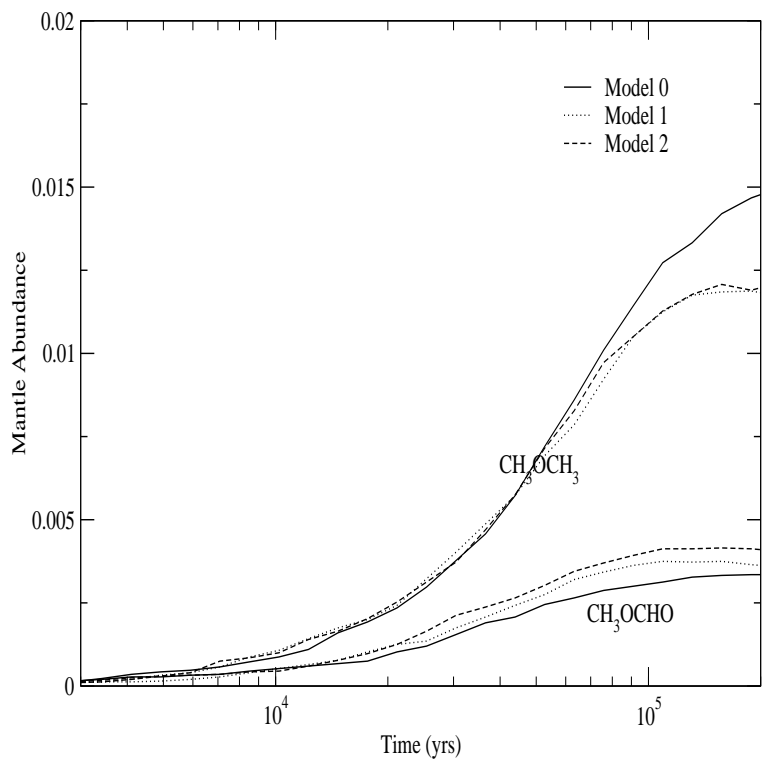}}
\caption{The abundance (in monolayers) of CH$_3$OCH$_3$ and CH$_3$CHO as a function of time for Models 0, 1, and 2 at 10 K}
\label{fig7}
\end{figure}

Figure~\ref{fig7} shows the abundances (in monolayers)  of two complex organic molecules, CH$_3$OCH$_3$ (dimethyl ether)  and CH$_3$OCHO (methyl formate) as a function of time for 
Models 0, 1, and 2 at 10 K, while Table~\ref{table5} lists their fractional abundances at $2 \times 10^{5}$ yr for both 10 K and 15 K.  All the individual CH$_3$OCH$_3$ and CH$_3$CHO molecules reside on normal sites.
We can see that despite the fact that there are many more radicals such as CH$_2$OH produced
in Models 1 and 2 than  in Model 0, the abundances of CH$_3$OCH$_3$ and CH$_3$CHO in all models are very similar, which shows that interstitial chemistry plays little role in their production.  
These complex organic molecules are produced on the top few monolayers, so that radicals deeply buried within the ice
do not contribute to their production at 10 K.  The formation of complex organic molecules at 10 K cannot occur by the radical-radical mechanism of \citet{Garrod2006} since heavy surface radicals in this model  do not move much until 30 K.
Instead, CH$_3$OCH$_3$  is mainly produced by the following mechanism.  
When an H atom hops to a site in the topmost monolayer  and combines with CH$_2$ to form CH$_3$, if there is a CH$_3$O radical below the newly formed CH$_3$, they
react immediately to form CH$_3$OCH$_3$. This is the same mechanism as that which  produces CO$_2$~\citep{Chang2012}.  
Similarly, when an O atom hops into a site where there is CH residing on top of CH$_3$O, O can combine with CH first and the newly
formed HCO can immediately react with CH$_3$O to form CH$_3$OCHO. Moreover, most of the CH$_3$OCH$_3$ and CH$_3$OCHO are produced
before $10^5$ yr, which shows that radicals such as CH or CH$_2$ are mainly produced by hydrogenation of atomic C, whose abundance
significantly drops in the gas phase with increasing time~\citep{Chang2012}. The amounts of dimethyl ether and methy formate produced by 10$^{5}$ yr on the ice are reasonably large but abundances of these species in cold cores can only be observed in the gas.


\begin{table}
\caption{Fractional Abundances of Granular Complex Organic Molecules  }
\label{table5}
\begin{tabular}{ccccccc}
  \hline
Species  &             Model 1      & Model 1        & Model 2          & Model 2     & Model 0      & Model 0    \\ 
              &              10 K        &    15 K            &       10 K          &  15 K               &    10 K     &  15 K  \\ \hline
CH$_3$OCH$_3$ & 1.2(-8)        & 1.8(-9)       & 1.2(-8)       & 1.1(-9)       &    1.5(-8)    & 1.2(-9) \\        
CH$_3$OCHO    & 3.6(-9)        & 1.0(-9)      & 4.1(-9)       & 9.2(-10)       &   3.4(-9)     & 9.0(-10) \\
\end{tabular}
\tablecomments{The fractional abundances are at a time of $2\times 10^5$~yr for 
Models 0, 1, 2 at different temperatures. a(-b) means a$\times 10^{-b}$.
}
\end{table}

\section{Comparisons with Previous Models and Observations}

The results reported in this paper are from the first simulation using a full gas-grain chemical reaction network with a microscopic Monte Carlo treatment of the grain mantle
chemistry.  It is interesting to compare these new results with our mantle results from a previous Monte Carlo model, in which we used a rather limited set of surface 29 surface reactions, and did not consider photodissociation or bulk diffusion~\citep{Chang2012}.  In the earlier paper, a comparison with older rate-equation models was also made.  Table \ref{table6} shows a comparison between our current results for CO, CO$_{2}$, CH$_{3}$OH, and CH$_{4}$ ices and those of a previous Monte Carlo model (Model 3) with the same parameters as our current Model 0 at 10 K so that only the number of grain mantle reactions differs.  We also make a thorough comparison with observational results of a compendium of sources in front of background stars, high-mass YSOs and low-mass YSOs \citep{Oberg2011},  adding NH$_{3}$ to the list of molecules.   The abundances are listed as percentages of the water ice abundance. The results from Models 0, 1, and 2 are very similar, 
so that bulk diffusion and photodissociation are clearly not critical for the species shown. 

The median observed water fractional abundance, about $5\times 10^{-5}$~\citep{Gibb2004, Pontoppidan2004,Boogert2004,Oberg2011}, 
 is in excellent agreement with all our current simulations  at $2 \times 10^{5}$ yr regardless of photon
penetration or temperature (10 K or 15 K). 
The agreement between current Model 0 (10 K) and the older Model 3 (10 K) is very good for CO, CO$_{2}$, and CH$_{3}$OH, while the older model predicts somewhat more methane.  Regarding observations, 
the results of our model simulations are in reasonable agreement with the summary of observational results of ~\citet{Oberg2011}, 
especially at 15 K, although our results for CO$_{2}$ are a factor of a few too low except for high mass stars, 
and our results for ammonia are high by a factor 3-5.  

\begin{deluxetable}{l c c c c c}
\tablecaption{Comparison of Observational and Ice Abundances By Different Models at $2 \times 10^{5}$ Yr}
\tablewidth{0pt}
\tablehead{       &  CO  & CO$_2$ &  CH$_3$OH & CH$_4$ & NH$_3$}
\startdata
  \hline
  Our New MC Models & & & & \\
  \hline
Model 1, 10 K     &           15.3 & 9.1 & 4.7 & 11.0 & 20.9 \\
Model 1, 15 K     &           30.8 & 11.3 & 3.8 & 5.8 & 16.1 \\
Model 2, 10 K     &           14.9 & 9.1  & 4.7 & 11.1 & 20.9 \\ 
Model 2, 15 K     &           30.5 & 11.3 & 3.8 & 5.3  & 15.9 \\ 
Model 0, 10 K     &           13.7 & 10.0 & 6.3 & 13.7 & 23.2 \\
Model 0, 15 K   &             29.6 & 12.2 & 4.7 & 6.1 & 17.2 \\
\hline
Our Previous MC Model & & & & \\
\hline
Old Model 3, 10 K   &     10.5 & 11.3 & 9.8 &21.2  & -- \\
\hline
Observations & & & &\\
\hline
Toward Low Mass Stars  &    29 & 29 & 3 & 5 &5 \\
Toward High Mass Stars &    13 & 13 & 4 & 2 & 5\\
Toward Background Stars &    31 & 38 & 4 &  & \\
\enddata
\label{table6}
\tablecomments{The abundances of  major ice mantle components are percentages with respect to the water ice.
  a(-b) means a$\times 10^{-b}$.}  
\end{deluxetable}


 In addition to total abundances, we can look at the abundances of major ice  species with respect to monolayer.
Observational results show that CO may exist in two distinct phases on grain surfaces formed at different times~\citep{Tielens1991,Oberg2011}, and a careful discussion with respect to our earlier Monte Carlo calculations can be found in~\citet{Chang2012}.
The polar phase, produced early in the mantle evolution, is H$_2$O-dominated, but also possesses 
abundant CO$_2$. The non-polar phase, produced later, is CO-dominated with some conversion to formaldehyde and methanol. 
Our simulations  $2\times 10^5$ yr do show a slight trend of more abundant CO and methanol at outer layers caused by the increased abundance of gas phase
CO and decreased abundance of atomic O  at later times.  However, even at the end of our simulation, 
H$_2$O is always the most abundant species in each layer, as can be seen in Figure~\ref{fig3}. This dominance cannot be explained by the ``short'' evolution in
our simulation because even with a three-phase rate equation model, which can be run to much longer times, H$_2$O is always  the most abundant species at each layer~\citep{Garrod2011}.
One possible explanation
is that grains can undergo some thermal processing, which induces segregation or distillation of CO$_2$ or CO~~\citep{Oberg2009c,Pontoppidan2008}. 
In our simulations,  CO molecules are able to move around within the ice mantle, but they cannot evaporate between 15 K and 20 K so they may be able to diffuse gradually to the top layers and form CO-rich layers.  
Thus,  the position of each molecule in the ice mantle may not be the actual positions where they were formed. 
However, a more complicated model with segregation or distillation coupled
with the full gas-grain network simulation is not the purpose of this work.

A major feature in the calculations reported here is the higher abundances of radicals with bulk diffusion and 
photon penetration included in Models 1 and 2. Moreover, it is photon penetration that generates 
bulk diffusing species deeply buried within the ice in our simulation.In the work by ~\citet{Garrod2013a}
photodissociation reactions are allowed to occur at equal rates within ice mantle,
however, the abundance of radicals was not reported in ~\citet{Garrod2013a}.  
A high abundance of radicals was obtained by photodissociation of bulk ice in experiments of Greenberg and colleagues \citep{Schutte1991} and shown to lead to explosions when warmed up.
Although the
GRAINOBLE model~\citep{Taquet2012}  also produces a significant abundance of radicals, the mechanism to form radicals is very different from that in our approach. 
In particular, the radicals formed  with the GRAINOBLE model simulation are actually normal species that react with other species, which is 
equivalent to that in our earlier work~\citep{Chang2007}, where we also produced many radicals.  However, with the introduction of
 the swap of H atoms and other species in the top 4 monolayers~\citep{Chang2012,Cuppen2009},    these radicals are almost
completely destroyed,  as occurs in Model 0. The radicals formed in Model 1 are, on the other hand, almost all formed by photodissociation of stable 
species deeply buried in the ice mantle and by reaction with photodissociation products. 
We can qualitatively compare our results with those of the molecular dynamics (MD) calculation by ~\citet{Andersson2008}. 
In this calculation, the products of 
photodissociated H$_2$O at the sixth monolayer, covered by 5 monolayers of water ice,
can recombine again with a probability  of 0.4 
so almost 60\% of the photodissociated H$_2$O molecules are frozen in the ice mantle in the form of OH radicals or water, formed from OH + H.  
This qualitatively agrees with our large abundance of 
radicals after introducing photon penetration.

The CH$_3$O radical, which is contained in our models, has been recently detected in the cold core B1-b with a low 
fractional abundance  of $4.7\times 10^{-12}$ in gas phase~~\citep{Cernicharo2012}.
Although the calculated granular CH$_3$O  fractional abundance  in Model 1 at 10 K is more than $10^{-8}$, because we do not
include any very efficient non-thermal desorption mechanisms, the gas-phase CH$_3$O radical abundance remains very low in our simulation. It is
not clear if we can produce the gas phase abundance of CH$_3$O radical with reactive desorption, which can efficiently
desorb surface species.     

Another new feature of our results is the prediction that complex organic molecules such as CH$_3$OCHO and CH$_3$OCH$_3$ can be formed on ice mantles
at temperatures as low as 10 K with fractional abundances of a few $\times$ 10$^{-8}$ for  CH$_3$OCH$_3$ and a few $\times$ 10$^{-9}$ for CH$_3$OCHO at $2\times 10^5$ yr. 
Recent observations ~\citep{Bacmann2012} show that the fractional abundance
of these two complex organic molecules in the gas phase in cold cores 
is about a few 10$^{-10}$. However, all CH$_3$OCHO or CHO$_3$CH$_3$ molecules in our models 
only exist on grain surfaces  because there is no efficient desoption mechanism included in our models. 
If 10\% of these two complex organic molecules can be desorbed  and  not destroyed in the gas phase, 
their gaseous fractional abundance  in our models will agree well with observations. 
However, because these two molecules can be destroyed in the gas phase,  it is not clear if we can
predict the observed abundance of these two complex organic molecules when an efficient desorption mechanism such as 
reactive desorption~\citep{Vasyunin2013a} is included.  

\section{Conclusions and Discussion}   
Despite the fact that the microscopic-macroscopic Monte Carlo method is exceedingly computationally demanding, 
we have successfully 
simulated the gas-grain chemistry of a cold dense core with a gas-grain reaction network containing $\approx$ 300 surface reactions.  
Our simulations, which follow the chemical evolution for $2 \times10^5$ yr,  represent a significant advance compared with our earlier
microscopic Monte Carlo simulations, with fewer than 30 surface reactions~\citep{Chang2012}.
The new simulations have been made possible by the use of an improved and more efficient algorithm, in which the species in the simulations are divided into three groups depending on the strength of their binding energy to the grain.  With this algorithm, we are able to follow additional processes involving the chemistry that occur in the bulk of the ice mantle. These processes include photon penetration and molecular dissociation caused by the penetrating photons. Moreover, 
we  are also able to follow a bulk chemistry occurring on interstitial sites located between sites of normal binding energy and maintained by photo-production of radicals. 
 We have focused the discussion in this paper on grain mantle species because species heavier than hydrogen and helium
cannot efficiently desorb at temperatures below 15 K.  The result is that the abundances of the
gas-phase species  in our three models are similar to those in two-phase models, where no distinction is made between surface and bulk mantle.
In a subsequent paper, we will include non-thermal reactive desorption so that all species on the grain surface can efficiently desorb; 
the abundances of gas phase species will then be more tightly coupled with the chemical composition of the ice mantle.

Our simulations show that the abundances of major stable ice species are not affected appreciably by photon penetration; thus, the simulation results concerning stable
major species in our previous study of cold cores are still valid.  Moreover, the results of both our past and current models agree reasonably well with observations of the abundance of
major species for sources toward high mass YSOs, low mass YSOs, and background stars. The major difference between the current work and past work has to do with the calculated granular abundances of radicals.   Formed mainly by photodissociation, the radicals can be enhanced in abundance by up to two orders of magnitude over models without this process in the bulk.  
The high abundance of radicals qualitatively 
agrees with recent studies using rate equations~\citep{Taquet2012} or molecular dynamical  simulations~\citep{Andersson2008}. 
The bulk diffusion of light species formed by photodissociation can also lead to the formation
of other radicals by interstitial reactions, a process  never before  included in previous models, although a similar model with pores has been reported \citep{Kalvans2013}.
With all the radicals  trapped within ice mantles,  especially at 10 K, chains of chemical reactions can occur when thermal 
processing of ice mantles occurs and the radicals are able to move more freely.  In this instance, explosions might even occur~\citep{Rawlings2013}.  This more complicated scenario will be included
in a future study.

 Our simulations also show that the chemistry occurring  on and in grain mantles
might be far more complex than it seems to be, especially when we treat ice mantle layers other than the topmost 
layer as a partially chemically active phase, and bulk diffusion of 
species within the ice mantle.  Indeed, the ice mantle, especially deep inside it,  
is a particularly ideal location for radical-radical association reactions because radicals will not find many other species with which to react. 
Other types of reactive species accreting from the gas will not approach deeply buried radicals, and in addition, there are few atoms available deep in the bulk. So the temperature range where radical-radical association reaction dominate should be much broader than
when these reactions happen on the topmost layers where radicals can be easily depleted by the large amount of species accreting from the gas phase
and, in addition, these radicals can evaporate when the temperature is high enough, as occurs in later stages.
Moreover, the fact that complex organic molecules such as CH$_3$OCH$_3$ can be formed on the topmost ice layers when the temperature is as low as
10 K further illustrates the major difference between reactions on the grain surface and those in the gas phase. Unlike the gas phase, 
because species are packed densely in ice mantles, three-body reactions such as the chain reaction mechanism  occur frequently within the ice.

It should be emphasized that interstitial bulk diffusion, which describes how molecules diffuse within a crystal lattice,
is a realistic physical process, and might be the dominant
bulk process at temperatures as low as 10 K. Moreover, when a molecule is photodissociated within the bulk, if potential minima around
this molecule are normally occupied, some products of the photodissociation reaction have to stay in interstitial sites. Thus, we must
take into account interstitial sites in modeling and simulation if we allow species deeply buried within ice mantle to be photodissociated.
However, it remains a large challenge to calculate the location of interstitial sites if we want to adopt a more rigorous off-lattice
microscopic Monte Carlo approach~\citep{Garrod2013b}. 
Based on such a simulation, the ice formed on a grain surface is non-porous if the gas density pertains to 
typical dark clouds ($n_H = 2\times 10^4~cm^{-3}$); however,
an ice mantle can be porous with a gas phase density much higher than that of typical 
dark clouds ($n_H \geq 2\times10^5~cm^{-3}$)~\citep{Garrod2013b}. 
If an ice mantle formed on a grain surface is
porous, our interstitial bulk diffusion model has to be modified.


 The details that can be obtained with a microscopic model of the grain chemistry should allow us to understand much better the different environments, both polar and non-polar, on grain mantles.  But to understand these environments better than we currently do,  we may have to add
more complex physical processes such as segregation, or distillation dynamics, into the simulation. 
These processes may change our results significantly 
because light species such as CO may tend to be more abundant in outer layers with segregation or distillation dynamics included
than in our current models. 
Although it would be worthwhile to study the further details of mantle chemistry, the unified microscopic-macroscopic Monte Carlo simulation has a serious problem:  it takes too much simulation time.
For instance, at 15 K, Models 1 and 2 take more than 8 days to simulate the evolution of the system through $2\times10^5$ years
with our Dell Presion T7500n workstation.
 Because of
the high computational cost, we have not yet been able to simulate the chemistry occurring during the warm-up phase of stellar evolution, which should be very interesting.   Besides this very serious problem, other issues remain outstanding.  
For example, most of the surface reactions in our network with experimentally determined activation energy barriers  have not been fitted by the competition mechanism in our simulation~\citep{HM2008}, with the exception of CO + H and H$_2$CO + H,
so it is not clear if we can still use these barriers, especially if determined from gas-phase reactions, for simulation by our approach.  
Finally,  whether we use a microscopic or even a macroscopic Monte Carlo simulation for the grain chemistry,  reactions involving H$_2$ on and in the grain mantle cannot be treated correctly because the rapid accretion 
and desorption of H$_2$ make Monte Carlo simulations currently not feasible. Some approximations and/or more powerful computers have to be introduced to solve this problem.

\acknowledgements

We are thankful to the referee for suggestions for revision that have improved this paper.  EH wishes to  acknowledge the support
of the National Science Foundation for his astrochemistry program. He also acknowledges support from the NASA Exobiology and Evolutionary Biology program through a subcontract from Rensselaer Polytechnic Institute.
We thank Anton I. Vasyunin for helpful discussions and allowing us to use his rate equation codes and network.


\begin{thebibliography}{12}

\bibitem[Akiyama et al.(1987)]{Akiyama1987}
Akiyama, A., Hosoi, T., Ishihara, I., Matsumoto, S., \& Niimi, T. 1987, IEEE Trans. on Comp. Aided Design CAD-6, 185

\bibitem[Andersson \& van Dishoeck(2008)]{Andersson2008} Andersson, S., \& van Dishoeck, E. F. 2008, A\&A, 491, 907

\bibitem[Bacmann et al.(2012)]{Bacmann2012} Bacmann, A., Taquet, V., Faure, A., Kahane, C., \& 
Ceccarelli, C. 2012, A\&A, 541, L12

\bibitem[{{Biham} {et al.}(2001) {Biham}, {Furman},{Pirronello}, \&{Vidali}}]{Biham2001}
{Biham}, O., {Furman}, I., {Pirronello}, V., \& {Vidali}, G. 2001, \apj, 553, 595

\bibitem[Boogert \& Ehrenfreund(2004)]{Boogert2004} Boogert, A. C. A., \& Ehrenfreund, P. 2004, in Astrophysics of Dust, ASP Conference Series, Vol. 309, ed. A. N.Witt, G. C. Clayton, B. T. Draine, p. 547


\bibitem[{{Caselli} {et al.}(1998){Caselli}, {Hasegawa}, \& {Herbst}}]{Caselli1998}
{Caselli}, P., {Hasegawa}, T.~I., \& {Herbst}, E. 1998, \apj, 495, 309

\bibitem[Cernicharo et al. (2012)]{Cernicharo2012}
Cernicharo, J., Marcelino, N., Roueff, E., et al. 2012, ApJL, 759, L43

\bibitem[{{Chang} \& {Herbst}(2012)}]{Chang2012}
{Chang}, Q., \& {Herbst}, E. 2012, \apj, 759, 147

\bibitem[{{Chang} {et al.}(2005) {Chang},{Cuppen},\& {Herbst} }]{Chang2005}
{Chang}, Q., {Cuppen}, H.~M., \& {Herbst}, E. 2005, \aap, 434, 599

\bibitem[{{Chang} {et al.}(2007) {Chang},{Cuppen},\& {Herbst} }]{Chang2007}
{Chang}, Q., {Cuppen}, H.~M., \& {Herbst}, E. 2007, \aap, 469, 973



\bibitem[{{Charnley}(1998)}]{Charnley1998}
{Charnley}, S.~B. 1998, \apj, 509, L121

\bibitem[{{Charnley} \& {Rodgers}(2009)}]{Charnley2009}
{Charnley}, S.~B., \& {Rodgers}, S.~D. 2009, in ASP Conf. Ser. 420, Bioastronomy 2007: Molecules, Microbes and Extraterrestrial Life,
ed. K. Meech et al. (San Francisco, CA: ASP), 29

\bibitem[{{Congiu} et al.(2012)}]{Congiu2012}
{Congiu}, E., et al. 2012, ApJ, 750: L12(1-4)


\bibitem[{{Cuppen} \& {Herbst}(2005)}]{Cuppen2005}
{Cuppen}, H.~M. \& {Herbst}, E. 2005, \mnras, 361, 565

\bibitem[{{Cuppen} {et~al.}(2009){Cuppen}, {van Dishoeck}, {Herbst}, \&
  {Tielens}}]{Cuppen2009}
{Cuppen}, H.~M., {van Dishoeck}, E.~F., {Herbst}, E., \& {Tielens}, A.~G.~G.~M.
  2009, \aap, 508, 275

\bibitem[{{Du} \& {Parise}(2011)}]{Du2011}
{Du}, F. \& {Parise}, B. 2011, \aap, 530, A131

\bibitem[{{Fayolle} {et al.}(2011)}]{Fayolle2011}
Fayolle, E. C., {\"O}berg, K., Cuppen, H. M., Visser, R., \& Linnartz, H. 2011, \aap, 529, A74


\bibitem[{{Fuchs} {et al.}(2009)}]{Fuchs2009}
Fuchs, G. W., Cuppen, H. M., Ioppolo, S., Romanzin, C., Bisschop, S. E., Andersson, S., van Dishoeck, E. F., \& Linnartz, H 2009, \aap, 505. 629

\bibitem[{{Garrod} (2008)}]{Garrod2008}
{Garrod}, R.  2008, \aap, 491, 239

\bibitem[Garrod(2013a)]{Garrod2013a}
Garrod, R. 2013a, \apj, 765, 60

\bibitem[Garrod(2013b)]{Garrod2013b}
Garrod, R. 2013b, \apj, 778, 158

\bibitem[{{Garrod} \& {Herbst}(2006)}]{Garrod2006}
{Garrod}, R., \& {Herbst}, E. 2006, \aap, 457, 927

\bibitem[{{Garrod} \& {Pauly} (2011)}]{Garrod2011}
{Garrod}, R., \& {Pauly}, T. 2011, \apj, 735, 15


\bibitem[{{Garrod} et al.(2009)}]{Garrod2009}
{Garrod}, R. T., {Vasyunin}, A. I., {Semenov}, D. A., {Wiebe}, D. S., \& {Henning}, Th. 2009, ApJ, 700, L43





\bibitem[Gibb et al.(2004)]{Gibb2004} Gibb, E. L., Whittet, D.C. B., Boogert, A. C. A., \& Tielens, A. G. G. M. 2004, ApJS, 151, 35

\bibitem[{{Gibson} \& {Bruck}(2000)}]{Gibson2000}
{Gibson}, M.~A., \& {Bruck}, J. 2000, J. Phys. Chem., 104, 1876

\bibitem[{{Gillespie}(1976)}]{Gillespie1976}
{Gillespie}, D.~T., 1976, J. Comp. Phys., 22, 403

\bibitem[{{Green et al.}(2001){Green},{Toniazzo},{Pilling},{Ruffle},{Bell},\& {Hartquist}}]{Green2001}
{Green}, N.~J.~B., {Toniazzo}, T., {Pilling}, M.~J., {Ruffle}, D.~P., {Bell}, N., \& {Hartquist} T.~W. 2001, \aap, 375, 1111


\bibitem[{{Hasegawa} \& {Herbst}(1993)}]{Hasegawa1993}
{Hasegawa}, T. I., \&  {Herbst}, E., 1993, \mnras, 263, 589

\bibitem[{{Hasegawa} {et  al.}(1992)}]{Hasegawa1992}
{Hasegawa}, T. I.,   {Herbst}, E., \& {Leung}, C.~M.1992, ApJS, 82, 167

\bibitem[{{Herbst} \& {Millar}(2008)}]{HM2008}
{Herbst}, E., \& {Millar}, T. J. 2008, in Low Temperatures and Cold Molecules, ed. I. W. M. Smith (London, Imperial College Press), p. 1


\bibitem[{{Herbst} \& {Shematovich}(2003)}]{Herbst2003}
{Herbst}, E., \& {Shematovich}, V.~I. 2003, Ap\&SS, 285, 725

\bibitem[Jenniskens et al(1995)]{Jenniskens1995}
Jenniskens, P., Blake, D. F., Wilson, M. A., \& Pohorille, A. 1995, \apj, 455, 389

\bibitem[{{Kalvans} \& {Shmeld}(2013)}]{Kalvans2013}
{Kalvans}, J., \& {Shmeld}, I. 2013, \aap, 554, 111

\bibitem[{{Lipshtat} \& {Biham}(2003)}]{Biham2003}
{Lipshtat}, A., \& {Biham}, O. 2003, \aap, 400, 585

\bibitem[{{{\"O}berg} {et al.}(2011a) {{\"O}berg}, {Boogert}, {Pontoppidan}, {Van Den Broek}, {van Dishoeck},
         {Bottinelli}, {Blake} \& {Evans II}}]{Oberg2011}
         {{\"O}berg}, K., {Boogert}, A.~C.~Adwin , {Pontoppidan}, K.~M.,  {Van Den Broek}, S.,  {van Dishoeck}, E.~F.,
         {Bottinelli}, S., {Blake}, G.~A., \&  {Evans II}, N. J., 2011a , \apj, 740, 109




\bibitem[{\"O}berg et al(2009a)]{Oberg2009a}
{\"O}berg, K., Fayolle, E. C., Cuppen, H. M., van Dishoeck, E.~F., \& Linnartz, H. 2009a, \aap, 505, 183



\bibitem[{{{\"O}berg} {et al.}(2007) {{\"O}berg}, {Fuchs}, {Guido}, {Award}, {Fraser},
           {Schlemmer}, {van Dishoeck} \& {Linnartz}}]{Oberg2007}
        {{\"O}berg}, K., {Fuchs}, G.~W., {Awad}, Z., {Fraser}, H.~J., {Schlemmer}, S., {van Dishoeck}, E.~F., \&
        {Linnartz}, H., 2007, \apj, 662, L23

\bibitem[{\"O}berg et al(2009b)]{Oberg2009b}
{\"O}berg, K., Linnartz, H., Visser, R., \&  van Dishoeck, E.~F. 2009b, \apj, 693, 1209



\bibitem[{\"O}berg {et al.}(2009c)]{Oberg2009c}
        {{\"O}berg}, K., {van Dishoeck}, E.~F., \&
        {Linnartz}, H., 2009c, \aap, 496, 281


\bibitem[{Pickles} \& {Williams}(1977)]{Pickles1977} Pickles, J. B., \& Williams, D. A. 1977, Ap\&SS, 52, 443

\bibitem[Pontoppidan et al.(2008)]{Pontoppidan2008} 
Pontoppidan, K. M., Boogert, A. c. A., Fraser, H. J., van Dishoeck, E. F., Balke, G. A., Lahuis, F., {\"O}berg, K., Evans, 
N. J., \& Salyk, C. 2008, \apj, 678, 1005 

\bibitem[Pontoppidan et al.(2004)]{Pontoppidan2004} Pontoppidan, K. M., van Dishoeck, E. F., \& Dartois, E. 2004, A\&A, 426, 925


\bibitem[Rawlings et al.(2013)]{Rawlings2013}
	Rawlings, J. M. C.,  Williams, D. A., Viti, S., Cecchi-Pestellini, C., Duley, W. W. 2013, \mnras, 430, 264

\bibitem[Schutte \& Greenberg(1991)]{Schutte1991} Schutte, W. A., \& Greenberg, J. M. 1991, A\&A, 244, 190

\bibitem[{{Semenov} {et al.}(2010){Semenov}, {Hersant}, {wakelam}, {Dutrey}, {Chapillon}, {Guilloteau}, {Henning},
          {Launhardt}, Pi\'etu, \& {Schreyer}}]{Semenov2010}
{Semenov}, D.,  {Hersant}, F., {Wakelam}, V., {Dutrey}, A., {Chapillon}, E., {Guilloteau}, S., {Henning}, T.
          {Launhardt}, R., Pi\'etu, V., \& {Schreyer} K. 2010, \aap, 522, A42

\bibitem[{{Shen} {et al.}(2004) {Shen}, {Greenberg}, {Schutte}, \& {van Dishoeck}}]{Shen2004}
{Shen}, C.~J., {Greenberg}, J.~M., {Schutte}, W.A., \& {van Dishoeck}, E.F. 2004, \aap, 415, 203

\bibitem[Smith et al.(2011)]{Smith2011} Smith, R. G., Charnley, S. B., Pendleton, Y. J., Wright, C. M., Maldoni, M. M., \& Robinson, G. 2011, \apj, 743, 131

\bibitem[{{Stantcheva} \& {Herbst}(2004)}]{Stantcheva2004}
{Stantcheva}, T., \& {Herbst}, E. 2004, \aap, 423, 241

\bibitem[{{Taquet} {et al.}(2012)}]{Taquet2012}
Taquet, V., Ceccarelli, C., \&  Kahane, C. 2012 \aap, 538, A42

\bibitem[{{Tielens} \& {Charnley}(1997)}]{Tielens1997}
{Tielens}, A.~G.~G.~M, \& {Charnley}, S.~B. 1997, Origins Life Evol. B., 27, 23

\bibitem[{Tielens et al.}(1991)]{Tielens1991} Tielens, A. G. G. M., Tokunaga, A. T.,  Geballe, T. R., \& Baas, F. 1991, \apj, 381, 181

\bibitem[{{Vasyunin} \& {Herbst}(2013a)}]{Vasyunin2013a}
Vasyunin, A.~I., \& {Herbst}, E. 2013a, \apj, 769, 34

\bibitem[{{Vasyunin} \& {Herbst}(2013b)}]{Vasyunin2013b}
Vasyunin, A.~I., \& {Herbst}, E. 2013b, \apj, 762, 86

\bibitem[{{Vasyunin} {et al.}(2009){Vasyunin},{Semenov},{Wiebe},\&{Henning}}]{Vasyunin2009}
{Vasyunin}, A.~I., {Semenov}, D.~A., {Wiebe}, D.~S., \& {Henning}, T. 2009, \apj, 691, 1459

\bibitem[{{Watanabe} \& {Kouchi}(2002)}]{Watanabe2002}
Watanabe, N., \& Kouchi, A.  2002, ApJ, 571, L173
\end{thebibliography}
\end{document}